\documentclass[12pt]{article}
\usepackage{bm}
\usepackage{amsmath}
\usepackage{amssymb}
\usepackage{xcolor}
\usepackage[mathcal]{euscript}
\usepackage{graphicx}
\usepackage{pgf}
\newcommand{\bfphi}{\mbox{\boldmath $\varphi$}}
\newcommand{\bfpsi}{\mbox{\boldmath $\psi$}}
\newcommand{\bfsigma}{\mbox{\boldmath $\sigma$}}

\begin{document}

\title{Lorentz Invariant Equations for Multiplets and Their Gauge}

\author{Mayer Humi\\
Department of Mathematical Sciences\\
Worcester Polytechnic Institute\\
100 Institute Road\\
Worcester, MA  01609\thanks {corresponding author, e-mail: mhumi@wpi.edu}}

\maketitle
\thispagestyle{empty}

\begin{abstract}
This paper explores the existence of kinematical gauge transformations 
for Lorentz invariant equations which describe a multiplet of two spin 
$\frac{1}{2}$ particles. For this multiplet the additional gauge invariance
can be in form of three different groups of transformation. 
This gauge is absent when the particles are treated separately. 
Some basic properties of the solutions for these "multiplet equations" 
are analyzed.
\end{abstract}

\newpage

\setcounter{equation}{0}

\section{Introduction}
The origin of gauge invariance relates to the Lorenz and Coulomb gauge 
for the electromagnetic potentials \cite{JJO,JJD}. 
The discovery of the inner symmetries for multiplets of elementary particles 
\cite{GEL} provided another wide range applications for gauge invariance 
\cite{BFF,JJD,LOR}.The objective of this paper is to 
demonstrate the existence of a kinematical gauge for Lorenz invariant 
equations for a multiplet of two spin $\frac{1}{2}$ particles \cite{YY}. 
It is conjectured that similar
results apply to Lorenz invariant equations describing larger multiplets. 
The existence of this kinematical gauge might be of interest within the 
context of the quark model \cite{DVJ,VVG} and nuclear matter which consist of 
multiplets of protons and neutrons.


Lorentz invariant equations describing ``elementary particles'' have
been the subject of many research papers and monographs
\cite{WIAG,IMGM,VBEW,HJB,HC,GVAW,MH1,MHSM,MH3}. (The first of these 
references contains an exhaustive list of citations on this subject). 
The most successful among such equations is the Dirac equation \cite{VMS} 
which describes simultaneously a particle of spin $\frac{1}{2}$ and its 
antiparticle.  For particles of higher spin the corresponding invariant 
equations might contain redundancies which must be eliminated by 
subsidiary conditions \cite{HJB,IMGM,WIAG}.

The application of this approach to the simultaneous description of
more than one particle is usually deemed to have a straightforward
answer in the form of the direct product of the equations and their
wave functions.

Another important argument that is made in this context is that of
ireducibility, i.e. an ``elementary particle'' must ``correspond'' to
an irreducible representation of the Lorentz group (otherwise it is not 
``elementary''). However such an argument does not preclude, at least in 
principle, the simultaneous description of two (or more) particles by one 
equation as is done by the Dirac equation.

In this paper we consider Lorentz invariant 8-dimensional systems of
equations which describe two spin $\frac{1}{2}$ particles of equal
mass and their antiparticles. We show that for these equations there exist  
groups of ``gauge transformation" \cite{LOR} which belong to one of the 
following groups $SO(3)$, $SO(2,1)$ and $E(2)$.

The plan of the paper is as follows:\,\,
In Section 2 we discuss Dirac equation and its possible "gauge".
In Sec. $3$ we study the construction of an 8 dimensional Lorentz invariant 
equations and derive the constraints and forms of their gauge.  
In Section $4$ we discuss plane wave solutions to these equations 
under different gauge transformations, multiplet interaction with 
an electromagnetic field and motion in a central force field.
We end in Section $5$ with conclusions.

\setcounter{equation}{0}

\section{Flashback on Dirac Equation}

Lorentz second order energy momentum invariant is
\begin{equation}
\label{1.1}
E^2-p^2-m^2=p^{\mu}p_{\mu}-m^2,\,\,\mu=0,1,2,3
\end{equation}
where $p^{\mu}$ is the 4-momentum $(E,p_x,p_y,p_z)$.
To factor this expression by first order terms Dirac attempted to rewrite it 
as
\begin{equation}
\label{1.2}
(\gamma^{\mu}p_{\mu}+m)(\gamma^{\nu}p_{\nu}-m)=
\gamma^{\mu}\gamma^{\nu}p_{\mu}p_{\nu}-m^2
\end{equation}
where $\gamma^{\mu}$ are $4\times 4$ matrices. To collapse this expression
to the one in (\ref{1.1}) one must have 
\begin{equation}
\label{1.3}
\gamma^{\mu}\gamma^{\nu} + \gamma^{\nu}\gamma^{\mu} =
\{\gamma^{\mu},\gamma^{\nu}\}=
2g^{\mu\nu},\;\;\;\;g^{\mu\nu} = \mbox{diag}(1,-1,-1,-1).
\end{equation}
Thus one obtains Dirac equation in the standard covariant notation
\begin{equation}
\label{1.4}
(\gamma^{\mu} p_{\mu}-mI_4)\psi = 0,\;\;\;\;\mu = 0,1,2,3
\end{equation}
where $I_k$ is the unit matrix in $k$ dimensions and
\begin{equation}
\label{1.5}
p_{\mu} = i\;\frac{\partial}{\partial x^\mu},\;\;\;x^\mu = (t,x,y,z)
\end{equation}
\begin{equation}
\label{1.6}
\gamma^0 = \left(\begin{array}{cr}
I_2  &0\\
0  &-I_2\\
\end{array} \right),\;\;\;\;\gamma^k = \left(\begin{array}{rr}
0   &\sigma_k\\
-\sigma_k   &0  \end{array}\right),\;\;\;k = 1,2,3.
\end{equation}
Here $\sigma_k$ are Pauli matrices
\begin{equation}
\label{1.7}
\sigma_1 = \left(\begin{array}{cr}
0  &1\\
1  &0\\
\end{array} \right),\;\;\;\;\sigma_2 = \left(\begin{array}{rr}
0   &-i\\
-i  &0  \end{array}\right),\;\;\; \sigma_3 = \left(\begin{array}{rr}
1   &0\\
0   &-1  \end{array}\right).
\end{equation}

However as a by product of this construction one obtains one additional
matrix
\begin{equation}
\label{1.8}
\gamma^5=i\gamma^0\gamma^1\gamma^2\gamma^3=\left(\begin{array}{cccc}
0  & 0 &1 & 0\\
0  & 0 &0 & 1\\
1  & 0 &0 & 0\\
0  & 1 &0 & 0 \\
\end{array}\right)
\end{equation}
which has the following properties
\begin{equation}
\label{1.9}
(\gamma^5)^2=I_4,\,\,\,\{\gamma^5,\gamma^{\mu}\}=0.
\end{equation}
This matrix has no direct relationship to \eqref{1.4} but in view of 
its existence and properties one is motivated to consider a possible
modification of the factorization in \eqref{1.2} in the form
\begin{equation}
\label{1.10}
(\gamma^{\mu}p_{\mu}+im_1\gamma^5+mI_4)(\gamma^{\nu}p_{\nu}+im_1\gamma^5-mI_4)
\end{equation}
where $m_1$ is a constant. Using the relations between the $\gamma$ matrices 
we find that the result of this multiplication yields 
$$
\gamma^{\mu}\gamma^{\nu}p_{\mu}p_{\nu}-(m_1^2+m^2).
$$
Thus we obtain the same right hand side as in \eqref{1.2} but with a mass 
shift in the particle mass. To see if this result can replicated by  
more general matrices $\beta=(b_{ij})$, we replaced $\gamma^5$ by $\beta$ 
in \eqref{1.10} where $m_1$ is a constant and $\beta$ is $4\times 4$ matrix. 
Expanding the left hand side of this expression we find that in order to recoup 
the right hand side of \eqref{1.10} (viz. nullifying the spurious terms in 
the multiplication) the matrix $\beta$ has to be a multiple of the 
matrix $\gamma^5$.

It follows then that Dirac equation does have a gauge in the form of the 
matrix $\gamma^5$ but this gauge affects only the mass of 
the particle represented by the equation.

\setcounter{equation}{0}

\section{Multiplet Equation}

A straightforward prescription to construct an equation describing two
spin $\frac{1}{2}$ particles with the same mass simultaneously is to
introduce
\begin{equation}
\label{2.1}
\Lambda^\mu = \left(\begin{array}{cc}
\gamma^\mu   &0\\
0  &\gamma^\mu  \end{array}\right)
\end{equation}
and consider the equation
\begin{equation}
\label{2.2}
(\Lambda^\mu p_\mu - mI_8)\psi = 0
\end{equation}
where  $\psi$ is made of two 4-dimensional spinors  $\psi^1, \psi^2$ viz.  
$\psi = \left(\begin{array}{c}  \psi^1\\ \psi^2
\end{array}\right)$.  
However in this form of the equation the components
$\psi^1, \psi^2$ are totally uncoupled and therefore there is no
advantage in the use of this simultaneous equation.

To explore a possible gauge in this representation of particle multiplets
we follow the "inspiration" derived from the the previous section and modify 
\eqref{2.2} by an additional term $im_1\beta_1$ where $\beta_1$ is an 
$8\times 8$ matrix and replace $I_8$ (in \eqref{2.2}) by $8\times 8$ matrix $\beta_2$. 
Thus we are seeking to represent the multiplet by an equation of the form
\begin{equation}
\label{2.3}
(\Lambda^\mu p_\mu+im_1\beta_1 - m\beta_2)\psi = 0
\end{equation}

To derive the constraints on the matrices $\beta_1$ and $\beta_2$
that are needed to satisfy \eqref{1.1} we multiply 
\begin{equation}
\label{2.4}
[(\Lambda^\mu p_\mu+im_1\beta_1+ m\beta_2)][(\Lambda^\mu p_\mu+im_1\beta_1-m\beta_2)\psi]
\end{equation}
Using the anti-commutation relations of the $\gamma$ matrices (which are 
inherited by the $\Lambda$ matrices) we find that in order to nullify  
the linear terms with $p^{\mu}$ in the resulting expression the matrices 
$\beta_1$ and $\beta_2$ have to satisfy
\begin{equation}
\label{2.5}
\{\beta_1,\Lambda^{\mu}\}=0,\,\,\,[\beta_2,\Lambda^{\mu}]=0.
\end{equation}
(where $\{\cdot,\cdot\}$ represents anti-commutation).

Using these relations yields the following reduced form for \eqref{2.4}
\begin{equation}
\label{2.6}
\left[-\left(\frac{\partial\psi}{\partial t}\right)^2+\left(\frac{\partial\psi}{\partial x}\right)^2+
\left(\frac{\partial\psi}{\partial y}\right)^2+\left(\frac{\partial\psi}{\partial z}\right)^2\right]I_8-[m^2\beta_2^2+m_1^2\beta_1^2+im m_1(\beta_2\beta_1-\beta_1\beta_2)]\psi
\end{equation}
Furthermore applying the constraints given by \eqref{2.5} on $\beta_1$ 
and $\beta_2$ we find that these matrices must be of the following form
\begin{equation}
\label{2.7}
\beta_1=\left(\begin{array}{cccccccc}
0   & 0 &b_1& 0  &0  &0  &b_2 &0 \\
0   & 0 &0  & b_1&0  &0  &0   &b_2 \\
b_1 & 0 &0  & 0  &b_2&0  &0   &0 \\
0   &b_1&0  & 0  &0  &b_2&0   &0 \\
0   & 0 &b_3& 0  &0  &0  &b_4 &0 \\
0   & 0 &0  & b_3&0  &0  &0   &b_4\\
b_3 & 0 &0  & 0  &b_4&0  &0   &0 \\
0   &b_3&0  & 0  &0  &b_4&0   &0 \\
\end{array}\right), \,\,\,\,
\beta_2=\left(\begin{array}{cccccccc}
c_1 & 0 &0  & 0  &c_2&0  &0   &0 \\
0   &c_1&0  & 0  &0  &c_2&0   &0 \\
0   & 0 &c_1& 0  &0  &0  &c_2 &0 \\
0   & 0 &0  &c_1 &0  &0  &0   &c_2 \\
c_3 & 0 &0  & 0  &c_4&0  &0   &0 \\
0   &c_3&0  & 0  &0  &c_4&0   &0 \\
0   & 0 &c_3& 0  &0  &0  &c_4 &0 \\
0   & 0 &0  & c_3&0  &0  &0   &c_4 \\
\end{array}\right).
\end{equation}

\subsection{Abelian Gauge}

One obvious choice of the gauge is to assume that the matrices $\beta_1$, 
$\beta_2$ in \eqref{2.6} commute. With this choice \eqref{2.6} corresponds
to a Klein-Gordon equation with a mass term $m^2\beta_2^2+m_1^2\beta_1^2$.
Applying the commutativity requirement to the matrices
$\beta_1$,$\beta_2$ in \eqref{2.7} yields the following system of three 
equations
\begin{equation}
\label{2.8}
b_2c_3-b_3c_2=0,\,\,\,(-c_1+c_4)b_3+c_3(b_1-b_4)=0,\,\,\,
(c_1 - c_4)b_2 - c_2(b_1 - b_4)=0 .
\end{equation}
This system has several solutions one of which is
$$
b_2=0,\,\,b_3=0,\,\,\,c_2=0,\,\,\,c_3=0
$$
while $b_1\,\,b_4,\,\,c_1,\,\,c_4$ remain as arbitrary parameters.
The mass matrix $M^2=m^2\beta_2^2+m_1^2\beta_1^2$ for this solution
is diagonal and we have
$$
M_{11}^2=M_{22}^2+M_{33}^2=M_{44}^2=c_1^2m^2+b_1^2m_1^2
$$
$$
M_{55}^2=M_{66}^2=M_{77}^2=M_{88}^2=c_4^2m^2+b_4^2m_1^2
$$
It follows that for this solution the gauge can separate the mass of the 
two particles. 

Other possible solutions of \eqref{2.8} with
$c_1 = c_4, c_2 = 0, c_3 = 0$ and $b_1,b_2,b_3,b_4,c_4$ as arbitrary
parameters leads to non-diagonal form of the matrix $M$.

\subsection{Nonabelian Gauge}

We now proceed to the case where the matrices $\beta_1$ and $\beta_2$ do not 
commute viz. $(\beta_2\beta_1-\beta_1\beta_2)\ne 0$. In 
this setting it is appropriate to rename $\beta_2$ as $J_3$ and $\beta_1$ 
as $J_1$  and then introduce an additional operator $J_2$ so that 
$\{J_1,\,J_2,\,J_3\}$ form  the Lie algebras of either $SO(3)$ or $SO(2,1)$.

We observe that the matrix representing $J_2$ do not enter explicitly 
in \eqref{2.4}. It follows that, in principle, the term representing this matrix
breaks the exact gauge. (In spite of the fact that the result of the 
multiplication in \eqref{2.4} satisfies a Klein-Gordon equation). However
if $m_1 \ll 1$ the gauge can be considered as approximate \cite{CGW,BFF,LOR}.

\subsubsection{SO(3) Gauge}

In this case the required commutation relations for the operators 
$\{J_1,\,J_2,\,J_3\}$  are
\begin{equation}
\label{2.9}
[J_1,\,J_2]=iJ_3,\,\,[J_2,J_3]=iJ_1,\,\,[J_3,J_1]=iJ_2
\end{equation}
As a first step towards satisfying these commutation relation we define
$J_2$ by the  
\begin{equation}
\label{2.10}
iJ_2= \beta_2\beta_1-\beta_1\beta_2=[J_3,J_1].
\end{equation}
Using \eqref{2.7} we obtain for $J_2$ the following form
\begin{equation}
\label{2.11}
J_2=i\left(\begin{array}{cccccccc}
0   & 0  &d_1 & 0  &0   &0   &d_3 &0 \\
0   & 0  &0   &d_1 &0   &0   &0   &d_3 \\
d_1 & 0  &0   & 0  &d_3 &0   &0   &0 \\
0   &d_1 &0   & 0  &0   &d_3 &0   &0 \\
0   & 0  &-d_2& 0  &0   &0   &-d_1&0 \\
0   & 0  &0   &-d_2&0   &0   &0   &-d_1\\
-d_2& 0  &0   & 0  &-d_1&0   &0   &0 \\
0   &-d2&0   & 0  &0   &-d_1&0   &0 \\
\end{array}\right)
\end{equation}
where
\begin{equation}
\label{2.12}
d_1=b_2c_3 - b_3c_2,\,\, d_2=(-c_1 + c_4)b_3 + c_3(b_1 - b_4),\,\,
d_3=(-c_1 + c_4)b_2 + c_2(b_1 - b_4)
\end{equation}
Applying this form of $J_2$ to the other two remaining commutation relations
in \eqref{2.9} and using \eqref{2.12} leads to a system of eight equations 
for the coefficients $b_i,c_i$, $i=1,2,3,4$. A solution of this system is
\begin{equation}
\label{2.13}
b_1 = c_2b_3,\, b_2 = -\frac{4b_3^2c_2^2-1}{4b_3},\, b_4 = -c_2b_3,\, c_1 = -1/2,\, c3 = 0,\, c4 = 1/2.
\end{equation}
with $b_3,\,c_2$ as parameters. With the values given by \eqref{2.12} and
\eqref{2.13} for the entries of of the operators $J_i$, $i=1,2,3$ it is 
straightforward to verify that these operators satisfy the commutation 
relations in \eqref{2.9}. 

The mass term in \eqref{2.6} is
\begin{equation}
\label{2.14}
M^2=m_1^2J_1^2+im m_1 J_2+m^2J_3^2
\end{equation}
The matrix $M^2$ is not Hermitian in general (viz. with arbitrary 
choice of the parameters $b_3,\,c_2$). However if we let $c_2=0$ and 
$b_3=\frac{1}{2}$ we obtain a Hermitian matrix of the following form:
\begin{equation}
\label{2.15}
M^2=\left(\begin{array}{cccccccc}
n   & 0  &0   & 0  &0  &0  &n_1 &0 \\
0   & n  &0   & 0  &0  &0  &0   &n_1 \\
0   & 0  &n   & 0  &n_1&0  &0   &0 \\
0   & 0  &0   & n  &0  &n_1&0   &0 \\
0   & 0  &-n_1& 0  &n  &0  &0   &0 \\
0   & 0  &0   &-n_1&0  &n  &0   &b_4\\
-n_1& 0  &0   & 0  &0  &0  &n   &0 \\
0   &-n_1&0   & 0  &0  &0  &0   &n \\
\end{array}\right)
\end{equation}
where $n=\frac{1}{4}(m^2+m_1^2)$ and $n_1=\frac{i}{2}m_1m$. Furthermore all
the eigenvalues of this matrix are $\frac{1}{4}(m^2+m_1^2)\pm \frac{1}{2}m_1m$.

For the choice of the parameters $b_3,\,c_2$ mentioned above we have the 
following representation for the matrices $J_i$, $i=1,2,3$
$$
J_3=diag(-1/2,-1/2,-1/2,-1/2,1/2,1/2,1/2,1/2)
$$
\begin{equation}
\label{2.17}
J_1=\left(\begin{array}{cccccccc}
0  &0  &0  &0  &0  &0  &1/2 &0 \\
0  &0  &0  &0  &0  &0  &0   &1/2 \\
0  &0  &0  &0  &1/2&0  &0   &0 \\
0  &0  &0  &0  &0  &1/2&0   &0 \\
0  &0  &1/2&0  &0  &0  &0   &0 \\
0  &0  &0  &1/2&0  &0  &0   &0\\
1/2&0  &0  &0  &0  &0  &0   &0 \\
0  &1/2&0  &0  &0  &0  &0   &0 \\
\end{array}\right),
\end{equation}
\begin{equation}
\label{2.18}
J_2=i\left(\begin{array}{cccccccc}
0   & 0  &0   & 0  &0  &0  &1/2 &0 \\
0   &0   &0   & 0  &0  &0  &0   &1/2 \\
0   & 0  &0   & 0  &1/2&0  &0   &0 \\
0   & 0  &0   & 0  &0  &1/2&0   &0 \\
0   & 0  &-1/2& 0  &0  &0  &0   &0 \\
0   &0   &0   &-1/2&0  &0  &0   &0 \\
-1/2& 0  &0   & 0  &0  &0  &0   &0 \\
0   &-1/2&0   &0   &0  &0  &0   &0 \\
\end{array}\right).
\end{equation}

We observe that for other values of $c_2$ and $b_3$ the multiplet equation is 
still PT symmetric \cite{BEN} and hence acceptable from a physical point of view.

\subsubsection{SO(2,1) Gauge}
In this case the required commutation relations for the operators
$\{J_1,\,J_2,\,J_3\}$  are
\begin{equation}
\label{2.19}
[J_1,\,J_2]=-iJ_3,\,\,[J_2,\,J_3]=-iJ_1,\,\,[J_3,\,J_1]=-iJ_2
\end{equation}
Using the commutation relation between $J_3,\, J_1$ and \eqref{2.7} we 
obtain the following form for $J_2$
\begin{equation}
\label{2.20}
J_2=i\left(\begin{array}{cccccccc}
0   & 0  &-d_1 & 0  &0   &0   &-d_2 &0 \\
0   & 0  &0   &-d_1 &0   &0   &0   &-d_2 \\
-d_1 & 0  &0   & 0  &-d_2 &0   &0   &0 \\
0   &-d_1 &0   & 0  &0   &-d_2 &0   &0 \\
0   & 0  &d_3& 0  &0   &0   &d_1&0 \\
0   & 0  &0   &d_3&0   &0   &0   &d_1\\
d_3& 0  &0   & 0  &d_1&0   &0   &0 \\
0   &d_3&0   & 0  &0   &d_1&0   &0 \\
\end{array}\right)
\end{equation}
where
\begin{equation}
\label{2.21}
d_1=b_2c_3 - b_3c_2,\,\,d_2=(-c_1 + c_4)b_2 + c_2(b_1 - b_4),\,\,
d_3=(-c_1 + c_4)b_3 + c_3(b_1 - b_4)
\end{equation}
Applying this form of $J_2$ to the other two remaining commutation relations
in \eqref{2.19} and using \eqref{2.21} leads to a system of eight equations
for the coefficients $b_i,c_i$, $i=1,2,3,4$. A solution of this system is
\begin{equation}
\label{2.22}
b_1 = c_2b_3,\, b_2 = -\frac{4b_3^2c_2^2-1}{4b_3},\, b_4 = -c_2b_3,\, c_1 = -1/2,\, c_3 = 0,\, c_4 = 1/2.
\end{equation}
with $b_3,\,c_2$ as parameters. Using the values given by \eqref{2.21} and
\eqref{2.22} for the entries of of the operators $J_i$, $i=1,2,3$ it is
straightforward to verify that these operators satisfy the commutation
relations in \eqref{2.19}.

The mass term in \eqref{2.6} is the same as in \eqref{2.14} and the resulting
matrix is not Hermitian in general (viz. with arbitrary
choice of the parameters $b_3,\,c_2$). However if we let $c_2=0$ and
$b_3=\frac{1}{2}$ we obtain a Hermitian matrix of the following form:
\begin{equation}
\label{2.23}
M^2=\left(\begin{array}{cccccccc}
n   & 0  &0   & 0  &0  &0  &-n_1 &0 \\
0   & n  &0   & 0  &0  &0  &0   &-n_1 \\
0   & 0  &n   & 0  &-n_1&0  &0   &0 \\
0   & 0  &0   & n  &0  &-n_1&0   &0 \\
0   & 0  &n_1& 0  &n  &0  &0   &0 \\
0   & 0  &0   &n_1&0  &n  &0   &0\\
n_1& 0  &0   & 0  &0  &0  &n   &0 \\
0   &n_1&0   & 0  &0  &0  &0   &n \\
\end{array}\right)
\end{equation}
where $n=\frac{1}{4}(m^2+m_1^2)$ and $n_1=\frac{i}{2}m_1m$. Furthermore
the eigenvalues of this matrix are $\frac{1}{4}(m^2+m_1^2)\pm \frac{1}{2}m_1m$.
For this choice of the parameters $b_3,\,c_2$ we have the following
representation for the matrices $J_i$, $i=1,2,3$
$$
J_3=diag(-1/2,-1/2,-1/2,-1/2,1/2,1/2,1/2,1/2)
$$
\begin{equation}
\label{2.24}
J_1=\left(\begin{array}{cccccccc}
0  &0  &0  &0  &0  &0  &1/2 &0 \\
0  &0  &0  &0  &0  &0  &0   &1/2 \\
0  &0  &0  &0  &1/2&0  &0   &0 \\
0  &0  &0  &0  &0  &1/2&0   &0 \\
0  &0  &1/2&0  &0  &0  &0   &0 \\
0  &0  &0  &1/2&0  &0  &0   &b_4\\
1/2&0  &0  &0  &0  &0  &0   &0 \\
0  &1/2&0  &0  &0  &0  &0   &0 \\
\end{array}\right),
\end{equation}
\begin{equation}
\label{2.25}
J_2=i\left(\begin{array}{cccccccc}
0   & 0  &0   & 0  &0  &0  &-1/2 &0 \\
0   &0   &0   & 0  &0  &0  &0   &-1/2 \\
0   & 0  &0   & 0  &-1/2&0  &0   &0 \\
0   & 0  &0   & 0  &0  &-1/2&0   &0 \\
0   & 0  &1/2& 0  &0  &0  &0   &0 \\
0   &0   &0   &1/2&0  &0  &0   &0 \\
1/2& 0  &0   & 0  &0  &0  &0   &0 \\
0   &1/2&0   &0   &0  &0  &0   &0 \\
\end{array}\right).
\end{equation}
We note also that each of these matrices is Hermitian.
\setcounter{equation}{0}
\section{Multiplet Solutions}

\subsection{Plane Wave Solutions}
For a free particle in the multiplet the wave function is the product of a 
plane wave and a spinor, $\bfphi(p^{\mu})$.
\begin{equation}
\label{4.1}
\bfpsi(x^{\mu})=\bfphi(p^{\mu})e^{i{\bf p}\cdot{\bf x}}
\end{equation}
Substituting this expression in \eqref{2.3} for a particle at rest where 
${\bf p}=0$ we obtain
\begin{equation}
\label{4.2}
(i\Lambda^0 \frac{\partial}{\partial t}+im_1\beta_1 - m\beta_2)\bfphi = 0
\end{equation}

Using the general form of the matrices $\beta_1$ and $\beta_2$
in \eqref{2.7} this leads to a system of eight equations for $\bfphi$
\begin{equation}
\label{4.3}
i\Lambda^0\frac{d\bfphi}{dt}=A\bfphi
\end{equation}
where
\begin{equation}
\label{4.4}
A=\left(\begin{array}{cccccccc}
-c_1m   & 0  &ib_1m_1   & 0  &-c_2m &0 &ib_2m_1 &0 \\
0   &-c_1m   &0   &ib_1m_1  &0 &-c_2m  &0 &ib_2m_1 \\
ib_1m_1   & 0  &-c_1m   & 0  &ib_2m_1&0  &-c_2m &0 \\
0   &ib_1m_1  &0   &-c_1m  &0  &ib_2m_1&0   &-c_2m \\
-c_3m   & 0  &ib_3m_1& 0  &-c_4m  &0  &ib_4m_1  &0 \\
0   &-c_3m   &0   &ib_3m_1&0  &-c_4m  &0   &ib_4m_1 \\
ib_3m_1& 0  &-c_3m   & 0  &ib_4m_1  &0  &-c_4m   &0 \\
0   &ib_3m_1&0   &-c_3m   &0  &ib_4m_1  &0   &-c_4m \\
\end{array}\right).
\end{equation} 
This system of equations can be solved now for the different possible gauges
with the proper values for the coefficients $b_i,\,c_i$, $i=1,\ldots,4$.

\begin{enumerate}

\item Abelian Gauge

For this gauge $b_2=b_3 =c_2= c_3 = 0$ while $b_1,\,b_4,\,c_1,\,c_4$ 
remain as arbitrary parameters. The general solution for $\bfphi$ is
\begin{equation}
\label{4.5}
\bfphi=\left(\begin{array}{c}
\frac{1}{2b_1m_1}[D_1(\alpha-c_1m)e^{i\alpha t}+
D_2(\alpha+c_1m)e^{-i\alpha t}] \\
\frac{1}{2b_1m_1}[D_3(\alpha-c_1m)e^{i\alpha t}+
D_4(\alpha+c_1m)e^{-i\alpha t}]  \\
\frac{1}{2}[-iD_1e^{i\alpha t}+iD_2e^{-i\alpha t}]  \\
\frac{1}{2}[-iD_3e^{i\alpha t}+iD_4e^{-i\alpha t}]   \\
\frac{1}{2b_4m_1}[D_5(\beta-c_4m)e^{i\beta t}+
D_6(\beta+c_4m)e^{-i\beta t}] \\
\frac{1}{2b_4m_1}[D_7(\beta-c_4m)e^{i\beta t}+
D_8(\beta+c_4m)e^{-i\beta t}]   \\
\frac{1}{2}[-iD_5e^{i\beta t}+iD_6e^{-i\beta t}]\\
\frac{1}{2}[-iD_7e^{i\beta t}+iD_8e^{-i\beta t}] \\
\end{array}\right).
\end{equation}
where $D_i$, $i=1\ldots 8$ are arbitrary constants and 
$$
\alpha=\sqrt{c_1^2m^2+b_1^2m_1^2},\,\,\,\beta=\sqrt{c_4^2m^2+b_4^2m_1^2}
$$

\item $SO(3)/SO(2,1)$ Gauge.

For a plane wave \eqref{4.2} contains only the matrices 
$\beta_1$ and $\beta_2$ which are the same for these two choices of the gauge.
Consequently the solutions for \eqref{4.3} are the same and we have:
\begin{equation}
\label{4.6}
\bfphi=\left(\begin{array}{c}
\left(-D_5{\mathrm e}^{-\frac{\mathit{m_1} t}{2}} +D_6{\mathrm e}^{\frac{\mathit{m_1} t}{2}} \right) {\mathrm e}^{\frac{i m t}{2}} \\
\left(-D_7{\mathrm e}^{-\frac{\mathit{m_1} t}{2}} +D_8{\mathrm e}^{\frac{\mathit{m_1} t}{2}} \right) {\mathrm e}^{\frac{i m t}{2}} \\
\left({D_2\mathrm e}^{-\frac{\mathit{m_1} t}{2}} -D_1{\mathrm e}^{\frac{\mathit{m_1} t}{2}} \right) {\mathrm e}^{-\frac{i m t}{2}} \\
\left(D_4{\mathrm e}^{-\frac{\mathit{m_1} t}{2}} -D_3{\mathrm e}^{\frac{\mathit{m_1} t}{2}} \right) {\mathrm e}^{-\frac{i m t}{2}} \\
\left(D_2{\mathrm e}^{-\frac{\mathit{m_1} t}{2}} +D_1{\mathrm e}^{\frac{\mathit{m_1} t}{2}} \right) {\mathrm e}^{-\frac{i m t}{2}} \\
\left(D_4{\mathrm e}^{-\frac{\mathit{m_1} t}{2}} +D_3{\mathrm e}^{\frac{\mathit{m_1} t}{2}} \right) {\mathrm e}^{-\frac{i m t}{2}} \\
\left(D_5{\mathrm e}^{-\frac{\mathit{m_1} t}{2}} +D_6{\mathrm e}^{\frac{\mathit{m_1} t}{2}} \right) {\mathrm e}^{\frac{i m t}{2}} \\
\left(D_7{\mathrm e}^{-\frac{\mathit{m_1} t}{2}} +D_8{\mathrm e}^{\frac{\mathit{m_1} t}{2}}\right) {\mathrm e}^{\frac{i m t}{2}} \\
\end{array}\right).
\end{equation}
Th
us we see that the gauge allows a ``mixing'' between the two
4-dimensional spinors in the same way that Dirac's equation allows a
``mixing'' among the 2-dimensional spinors.
\end{enumerate}
\subsection{Electromagnetic Coupling}

Using minimal coupling with an electromagnetic potential 
${\bf A}=(\phi, A_1,A_2,A_3)$
we obtain from \eqref{2.3} the following equation for  $\psi$.
\begin{equation}
\label{4.13}
[\Lambda^{\mu}(i\partial_{\mu}-qA_{\mu})+im_1\beta_1 -m\beta_2]\psi=0
\end{equation}
where  $q$ is the electric charge.
Multiplying this equation from the left by
$$ 
[\Lambda^{\mu}(i\partial_{\mu}-qA_{\mu})+im_1\beta_1 +m\beta_2]
$$
we obtain after a long algebra (and using eq. (\ref{2.9})) that
\begin{eqnarray}
\label{4.14}
&&\left\{(p_{\mu}-qA_{\mu})^2+iq\Lambda^0
\left[\Lambda^k\left(\frac{\partial A_0}{\partial x_k}+\frac{\partial A_k}{\partial t}\right)\right]+ 
\Lambda^3\Lambda^2\left[\frac{\partial A_3}{\partial y}-\frac{\partial A_2}{\partial z}\right]\right\}\psi+ \\ \notag
&&\left\{\Lambda^1\Lambda^3\left[\frac{\partial A_3}{\partial x}-\frac{\partial A_1}{\partial z}\right]+
\Lambda^2\Lambda^1\left[\frac{\partial A_2}{\partial x}-\frac{\partial A_1}{\partial y}\right] 
-[m^2\beta_2^2 +im_1m(\beta_1\beta_2 - \beta_1\beta_2)]\right\}\psi=0
\end{eqnarray}
This can be rewritten as
\begin{eqnarray}
\label{4.15} 
&&\left[(p_{\mu}-qA_{\mu})^2-iq\Lambda^{0}\Lambda^kE_k+\Lambda^3\Lambda^2B_1+
\Lambda^1\Lambda^3B_2+\Lambda^2\Lambda^1B_3\right]\psi \\ \notag
&&-\left[m_1^2\beta_1^2+ m^2\beta_2^2 +im_1m(\beta_1\beta_2 - 
\beta_1\beta_2)\right]\psi=0
\end{eqnarray}
where
$$
{\bf E}=-{\bf \bigtriangledown}A_0-\frac{d{\bf A}}{dt}=
-{\bf \bigtriangledown}\phi-\frac{d{\bf A}}{dt}
$$
$$
{\bf B}={\bf \bigtriangledown \times} {\bf A}
$$
(in this equations ${\bf A}=(A_1,A_2,A_3)$)

This result shows that the impact of the different gauges on the response of 
the particles to an $E-M$ field. 

\subsection{Motion in a Central Force Field with Abelian Gauge}

In this section we consider motion of a particle in a central force
field due to a multiplet of two spin $1/2$ particles with Abelian
gauge. To this end we shall let the entries of the matrices 
$\beta_1$ and $\beta_2$ in \eqref{2.7} satisfy
$$
b_2 = 0, b_3 = 0, c_2 = 0, c_3 = 0.
$$
Furthermore to simplify this treatment we shall assume that
$c_1=c_4$ and $b_1=b_4$.
 
Rewriting  \eqref{4.13} for $\psi$ in terms of four 2-d spinors  
$\phi_1\ldots\phi_4$ we obtain the following equations for the 
motion of a particle in a central force field whose potential is $V(r)$.
\begin{equation}
\label{5.10}
(i\frac{\partial}{\partial t}-V)\phi_1+(\bfsigma\cdot{\bf p})\phi_2-
c_1m\phi_1+ib_1m_1\phi_2=0,
\end{equation}
\begin{equation}
\label{5.11} 
(i\frac{\partial}{\partial t}-V)\phi_2+(\bfsigma\cdot{\bf p})\phi_1+
c_1m\phi_2-ib_1m_1\phi_1=0,
\end{equation}
\begin{equation}
\label{5.12} 
(i\frac{\partial}{\partial t}-V)\phi_3+(\bfsigma\cdot{\bf p})\phi_4-
c_1m\phi_3+ib_1m_1\phi_4=0,
\end{equation}
\begin{equation}
\label{5.13} 
(i\frac{\partial}{\partial t}-V)\phi_4+(\bfsigma\cdot{\bf p})\phi_3+
c_1m\phi_4-ib_1m_1\phi_3=0,
\end{equation}
where $\bfsigma=(\sigma_1,\sigma_2,\sigma_3)$.

Subtracting \eqref{5.12} from \eqref{5.10} and \eqref{5.13} from
\eqref{5.11} we obtain;
\begin{equation}
\label{5.14}
(i\frac{\partial}{\partial t}-V)(\phi_1-\phi_3)r+
({\bfsigma} \cdot {\bf p})(\phi_2-\phi_4)-c_1m(\phi_1-\phi_3) = 0\\
\end{equation}
\begin{equation}
\label{5.15}
(i\frac{\partial}{\partial t}-V)(\phi_2-\phi_4)+ 
({\bfsigma} \cdot {\bf p})(\phi_1-\phi_3)-c_1m(\phi_2-\phi_4)= 0
\end{equation}
Eqs. (\ref{5.14}), (\ref{5.15}) for  $\phi_1-\phi_3, \phi_2 - \phi_4$
are exactly the same as for a Dirac particle in a central force field
whose solutions are well known (see e.g. \cite{LIS}).
On the other hand if we use \eqref{5.10}-\eqref{5.11} to solve for 
$\phi_1,\,\phi_2$ (or \eqref{5.12}-\eqref{5.13} for 
$\phi_3,\,\phi_4$) the solutions will depend on the gauge if
$b_1$ is not zero. 

\section{Conclusions}

This paper considered a multiplet of two spin $\frac{1}{2}$ particles 
with equal (or almost equal mass). It demonstrated that for a Lorentz 
invariant equation describing this multiplet there exists a kinematical 
gauge of three types. The rest of the paper explored briefly    
the possible impact of this gauge on the interaction of this multiplet
with an external electromagnetic field and its motion
in central force field.


As an extension of these results it might be appropriate 
to examine the existence (and types) of this kinematical gauge for 
larger multiplets of particles and derive applications of this gauge 
to these multiplets.



\vspace*{.25in}

\begin{thebibliography}{www}

\bibitem{BEN} C. M. Bender, D.W. Hook, "PT-symmetric quantum mechanics"
Rev. Mod. Phys. {\bf 96} 045002 (2024) DOI: 10.1103/RevModPhys.96.045002

\bibitem{BFF} Berghofer P, François J, Friederich S, Henrique Gomes, 
Guy Hetzroni, Axel Maas and René Sondenheimer,"Gauge Symmetries, 
Symmetry Breaking, and Gauge-Invariant Approaches". 
Cambridge University Press; (2023). 

\bibitem{VBEW} V. Bargman and E.P. Wigner - Proc. Nat. Acad. Sci US
{\bf 34} p. 221 (1948).

\bibitem{HJB} H. J. Bhabha - Rev. Mod Phys. {\bf 21} p. 451 (1949)

\bibitem{CGW} W. Caudy, J. Greensite, "Ambiguity of spontaneously broken gauge
symmetry". Physical Review D, {\bf 78}(2), 025018.(2008)

\bibitem{DVJ} G. Ding, J.W.F. Valle, "The symmetry approach to quark and lepton 
masses and mixing" Physics Reports 1109, 1–105 (2025)

\bibitem{WIAG} W.I. Fushchich and A.G. Nikitin, "Symmetries of
Equations of Quantum Mechanics", Allerton Press, NY.(1994)

\bibitem{VVG} V.V. Gligorov, "Quark flavor physics: Status and future 
prospects", Inter. J. of Modern Physics A, {\bf 38} p.2330 (2023)

\bibitem{HC}  Harish-Chandra - Phys. Rev. {\bf 71} p. 793 (1947)

\bibitem{IMGM}  I.M.Gelfand, R.A. Minlos and Z. Ya Shapiro -
"Representations of the rotation and Lorentz groups and their
applications", Pergamon Press, NY.(1963)

\bibitem{GEL} M. Gell-Mann, Y. Ne'eman, eds. The Eightfold Way, 
W. A. Benjamin (1964)

\bibitem{MH1} M. Humi - J. Math. Phys. {\bf 10} p. 1877: ibid p. 1900.(1969)

\bibitem{MHSM} M. Humi, S. Malin - Phys. Rev. {\bf 187} p. 2278, (1969).

\bibitem{MH3} M. Humi - J. Math. Phys. {\bf 11} p. 222 (1970).

\bibitem {JJO} J.D. Jackson and L.B. Okun, "Historical roots of gauge 
invariance", Reviews of Modern Physics, {\bf 73} (3) pp. 663–680 (2001).

\bibitem{JJD} J.D. Jackson,"From Lorenz to Coulomb and other explicit gauge 
transformations". Am. J. Phys. 70 (9): 917–928 (2002).

\bibitem{LOR} L. O'Raifeartaigh - "The Dawning of Gauge Theory",
Princeton Univ. Press, Princeton, NJ. (1997).

\bibitem{LIS} L. I. Schiff - Quantum Mechanics, McGraw Hill, NY (1968).

\bibitem{VMS} V.M. Simulik - "The Dirac equation near centenary: a contemporary 
introduction to the Dirac equation consideration",
J. Phys. A: Math. Theor. {\bf 58}, 053001 (2025).

\bibitem{GVAW} G. Velo and A.S. Wightman (eds.) - Invariant wave
equations, Springer Verlag (1978), NY. (1978).



\bibitem{YY} Y. Yaremko - "On the validity of Lorentz Dirac equations",
  J. Phys. A {\bf 35} p. 831 (2002).

\end{thebibliography}
\end{document}